\documentclass[useAMS,usenatbib]{mn2e}

\usepackage{psfig}
\usepackage[dvips]{graphicx}
\usepackage{lscape}

\def\lesssim{\mathrel{\hbox{\rlap{\hbox{\lower4pt\hbox{$\sim$}}}\hbox{$<$}}}}
\def\gtrsim{\mathrel{\hbox{\rlap{\hbox{\lower4pt\hbox{$\sim$}}}\hbox{$>$}}}}

\title{On the universality of void density profiles}
\author[Ricciardelli et al.]{E. Ricciardelli$^{1}$\thanks{E-mail:
    elena.ricciardelli@uv.es}, V. Quilis$^{1}$, J. Varela$^{2}$\\
$^{1}$Departament d'Astronomia i Astrofisica, Universitat de Valencia, c/ Dr. Moliner 50, E-46100 - Burjassot, Valencia, Spain\\
$^{2}$Centro de Estudios de F\'{\i}sica del Cosmos de Arag\'on (CEFCA), Plaza San Juan 1, 44001 Teruel, Spain
}

\begin{document}
\date{Accepted ...  Received ...; in original form ...
}
%\pagerange{\pageref{firstpage}--\pageref{lastpage}} \pubyear{...}
\maketitle
\label{firstpage}
\begin{abstract}

   The massive exploitation of cosmic voids for precision cosmology in
the upcoming dark energy experiments, requires a  
robust understanding of their internal structure, particularly of their
density profile.
We show that the void density profile is insensitive to the void radius both 
in a catalogue of observed voids and in voids from a large cosmological 
simulation.
However, the observed and simulated voids display remarkably different
profile shapes, with the former having 
much steeper profiles than the latter. 
Sparsity can not be the main reason for this discrepancy, as we
demonstrate that the profile can be recovered with
reasonable accuracy even with
very sparse samples of tracers. 
On the other hand,  the observed profile shows a significant dependence
on the galaxy sample used to trace the matter distribution. 
Samples including low-mass galaxies lead to shallower profiles with
respect to the samples where only massive galaxies are used, as faint
galaxies live closer to the void centre.  We argue that
galaxies are biased tracers when used to probe the matter distribution
within voids.

\end{abstract}
 

\section{Introduction}\label{intro}

Large redshift surveys \citep{York00, Colless01}  and
cosmological simulations \citep{Bond96} have
revealed that galaxies are distributed inside a cosmic web 
of walls, filaments and compact clusters. Such a web encloses large
underdense regions, referred to as cosmic voids. 

Voids were first recognized in the earliest redshift surveys
\citep{Gregory78, Kirshner81}  as huge empty holes in the galaxy distribution. 
Nowadays, there is a general consensus in that voids occupy most of the
volume of the Universe \citep{Sheth04, Vandew11, Pan12},
although there is  not yet
an agreement on how a genuine void should be defined. Several void finders, which are  based on different
principles, have been developed. Voids can be identified as
spherical regions devoid of galaxies/haloes  \citep{Gottlober03,
  Patiri06b, Varela12} or underdense regions, relying on the continuous
density field \citep{Plionis02, Colberg05}. More complex
algorithms  able to capture the complex morphology of voids also exist 
\citep{Platen07, Neyrinck08, AragonCalvo13}.  Despite their
different definition of voids, all these
void finders agree in that voids ere extremely empty in the centre and
 show a sharp increase in the density towards the voids edges (e.g. \citealt{Colberg08}).

Voids are believed to originate from negative density fluctuations in the primordial
density field. As a result of their underdensity, they are subject to 
an effective repulsive peculiar gravity, causing their expansion. 
As a consequence of
such an expansion, the matter within the voids evacuates from the interior
and accumulates to the boundaries. This leads to void density profiles
that evolve towards a reverse top-hat shape \citep{Sheth04}.

A considerable appeal of cosmic voids is their potential in
probing cosmological parameters. In particular, being almost devoid of
matter, they are extremely sensitive to the nature of
dark energy.  Indeed, the void ellipticity and its evolution through cosmic
time are intimately connected with the local tidal tensor, which, in
turn, depends on the dark energy content  \citep{Park07, Lavaux10, Bos12}.
Voids  are also the ideal candidate for probing the
expansion history of the Universe through the Alcock-Paczynski test
\citep{AP79}, using the average shape of stacked
voids \citep{Lavaux12, Sutter12a}. The application
of such a test to the voids that will be identified in the future Euclid survey \citep{Laureijs11} promises to
outperform Baryonic Acoustic Oscillation by an order of magnitude in
accuracy. 

The huge potentiality of voids for precision cosmology requires a
robust knowledge of their internal structure, particularly of the 
density profiles. 
Works based on cosmological simulations  \citep{Colberg05, 
Ricciardelli13} indicate that the void density profile is
universal. As such,  it does not depend on void size.
On the observational side, the ideal approach to directly constrain
the void density profile is through the  weak lensing signal of
stacked voids \citep{Krause13}. However, the number of voids available from
spectroscopic catalogues is still limited to provide a robust
measurement of the signal \citep{Melchior13}.
At present, we can 
only rely on the galaxy distribution to trace the density within voids
\citep{Sutter12b}. Thus, to robustly assess a void model to describe the universal
density profile, one also needs to   assess the systematic effects arising
from the use of the sparse galaxy sampling.

In a previous work (\citealt{Ricciardelli13}, hereafter RQP13), 
we have shown, by means of a cosmological simulation, that a two parameters law can be used to fit
the density profile of voids of any size, density, morphology and
redshift.  The best-fit parameters show some dependence on
redshift, density, and, on a less degree, morphology, but they are
almost independent on the void size, although the limited statistics
prevented us to draw robust conclusions.  In this work, we want to test
this model and its dependence on void radius, 
against an observed catalogue of
voids and a larger simulation, thus dramatically increasing the
statistics. In doing so, we provide a robust determination of the
systematic effects arising when using the sparse distribution of
void galaxies as  density
tracers. 

The structure of the paper is as follows. In Section \ref{sdss} we
introduce our catalogue of observed voids, in Section \ref{sim} we
describe the simulation used and our void identification
procedure. The results on the void density profiles are discussed in
Section \ref{profiles}. We conclude in Section \ref{conclu}.

\section{The SDSS void catalogue}\label{sdss}

The catalogue of cosmic voids used for the present analysis has been
described in \citet{Varela12}. Here we only give a brief
description of  the main
features and the changes with respect to that work.  

The galaxy sample used for void
identification has been extracted from the New York
University Value-Added Galaxy Catalog\footnote{http://sdss.physics.nyu.edu/vagc/} (NYU-VACG; \citealt{Blanton05}), based on the
photometric and spectroscopic catalog of
SDSS/DR7\footnote{http://cas.sdss.org/astrodr7/en}, complete down to
$r\sim 17.8$.  
These authors also provide stellar masses computed with the code
  kcorrect (version 4.1.4) following the prescriptions of
  \citet{Blanton07}. Stellar masses have been computed assuming $h=1$.
To guarantee the
homogeneity of the sample and avoid the detection of spurious voids, a
complete catalogue up to redshift $0.12$ and down to magnitude
$M_r-5logh=-20.17$ has been used. Using this galaxy sample, voids are
defined as spherical regions devoid of galaxies. 
In the original catalogue of \citet{Varela12} only voids larger
than $10 \, h^{-1}\,  Mpc$  were considered for the analysis. 
In this work, in order to increase the number of voids, we have
extended the original catalogue to include voids down to $7 \,
h^{-1}\,  Mpc$. Moreover, we relax the assumption on void overlapping of
\citet{Varela12} and consider as separate voids all overlapping
voids whose distance between centers is larger than the radius of the
largest void. 
The final catalogue contains $4453$ voids, with radius as large
as $18.7 \, h^{-1}\, Mpc$. We find a total of $44617$ void
galaxies, which are, by definition, fainter than  $M_r-5logh=-20.17$.

\begin{figure}
\includegraphics[width=\columnwidth]{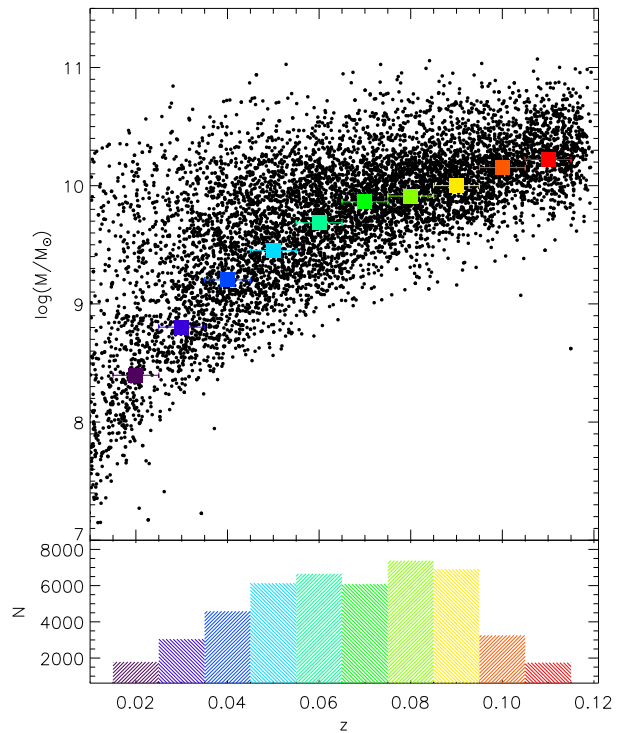}
\caption{Upper panel: void galaxies (black points) in the redshift stellar mass
  plane. For the sake of clearness only a randomly selected subsample, including
  15\% of the galaxies, has been plotted. The coloured points
  indicate the stellar mass threshold adopted for each redshift
  bin. Lower panel: number of galaxies with redshift less than $z$ and
  more massive than the threshold mass at $z$. }
\label{completeness} 
\end{figure}

Figure \ref{completeness} shows how our void galaxies populate the
redshift stellar mass plane. For each redshift bin we compute 
a stellar mass threshold (coloured points), above which the sample can
be considered complete. 
To choose this mass, we have computed the number counts in mass bins
and considered the threshold mass as the central mass of the bin having the largest counts.

\section{Voids in simulations}\label{sim}

The simulation used in this work has been performed with the
hydrodynamical code  MASCLET 
(Quilis 2004). MASCLET couples an Eulerian approach for describing the gaseous
component with an N-body scheme for treating the dark-matter, collisionless
component.  Gas and dark matter  are coupled by the gravity solver. To
gain spatial and temporal resolution an adaptive mesh refinement (AMR) 
scheme is implemented. 

The numerical  simulation was run  assuming a spatially  flat $\Lambda
CDM$  cosmology, with  the following  cosmological  parameters: matter
density    parameter,    $\Omega_m=0.27$;    cosmological    constant,
$\Omega_{\Lambda}=\Lambda/{3H_o^2}=0.73$;  baryon  density  parameter,
$\Omega_b=0.045$;  reduced Hubble  constant, $h=H_o/100  km\, s^{-1}\,
Mpc^{-1}=0.71$;  power  spectrum index,  $n_s=1$;  and power  spectrum
normalisation, $\sigma_8=0.8$.

The initial conditions are set up at z = 100, using a CDM transfer
function from \citet{EiHu98}, for a cube of comoving side
length $512 \, h^{-1}\,  Mpc$.  
The computational domain is discretized with $512^3$
cubical cells. The mass resolution is thus $\sim6\times
10^{10}\,h^{-1}\,M_{\odot}$  and the coarse spatial resolution is $1 \, h^{-1}\,  Mpc$.

Following the philosophy of the simulation presented in RQP13, we have designed the simulation to follow the formation
and evolution of low-density regions. 
Contrary to the common practice in AMR simulations, where the high
density regions are refined, we use more resolution in low density
regions.
During the evolution, the regions in the coarse grid are refined
based on the local density, when  $\rho/\rho_B<10$, being $\rho$ 
and $\rho_B$ the total density and the background density, respectively.
The ratio  between the cell  sizes for a  given level ($l+1$)  and its
parent   level  ($l$)   is,   in  our   AMR  implementation,   $\Delta
x_{l+1}/\Delta  x_{l}=1/2$.  
Since in this work we are not interested in the study of
the void sub-structures and void galaxies, in this simulation we have
only used  one level of refinement.  The best spatial resolution is therefore $0.5 \, h^{-1}\,  Mpc$.

Voids are identified in the simulated volume using the void finder
algorithm described in RQP13. This algorithm
relies on the continuous density field (including dark matter and gas)
to identify the low density regions, that we define as voids. It is
based on the basic assumptions that the velocity divergence of the gas
within the void is always positive (as a result of void expansion) and
that the density at the void edges has a sharp increase.
Broadly speaking, it performs the following steps. It first marks
cells as candidate for being centers of voids when their overdensity
 is below a threshold limit and the velocity
divergence is positive. It then expands these volumes by adding
cells on each coordinate directions until one of the conditions that
define the void edge is reached. Void edges are reached when the
velocity divergence becomes negative or the density gradient exceeds a
threshold value. The  procedure thus provides the protovoid,  the minimum
rectangle parallelepiped contained within a void. To build the actual
void, protovoids are allowed to merge with each other when the
ratio between the overlapping volume and the largest void is within
$0.5$ and $0.6$. The free parameters involved in the procedure have
been set by means of extensive tests of the code on a set of
Montecarlo mock voids, as well as on the voids in the simulation.  We
adopt the same reference values as in RQP13. The
density and the velocity divergence used are those defined in the base
level (l=0) grid.

The final sample of simulated voids includes a total of $\sim 35000$ voids,
filling 60\% of the simulated volume and with typical overdensity $\rho/\rho_B=0.2$.
For the analysis of the density profiles, we restrict the sample only
to large voids, with  effective radius\footnote{The effective
  radius is defined as the radius of the sphere having the same volume
  of the void. } $R_e>7\,h^{-1}\,Mpc$.
We also exclude voids having too large porosity and ellipticity, as 
they are the most affected by contamination from non-void regions. We end
up with 3186 voids.

\section{Void density profiles}\label{profiles}

\begin{figure}
\includegraphics[width=\columnwidth]{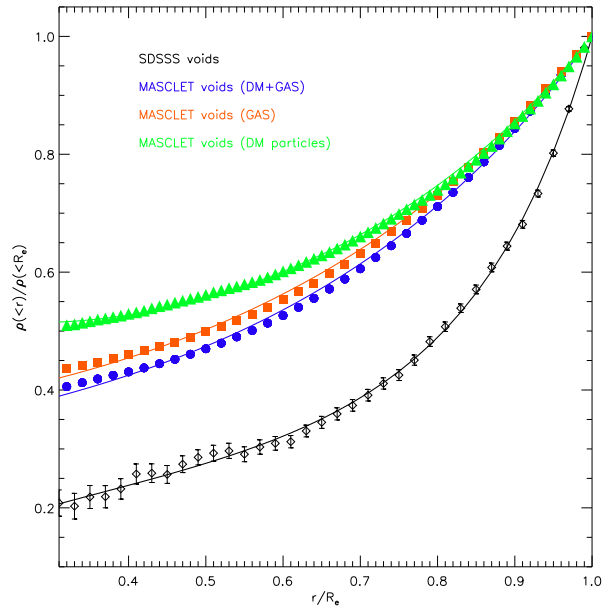}
\caption{Stacked density profiles for all voids larger than $7\,h^{-1}\, Mpc$ identified in the SDSS
  database (black diamonds) and the best-fit model (black line). As a
  comparison, the coloured symbols indicate the mean profiles of MASCLET
  voids larger than $7\,h^{-1}\, Mpc$.  The simulated profiles are
  computed using various density tracers: total - dark matter plus
  gas - density field (blue circles), gas density field (orange
  squares) and dark matter particles (green triangles). 
  The solid coloured lines indicate the best-fits for each curve. 
  $R_e$ refers to the radius of the voids. For the simulated
voids, which have arbitrary shape, $R_e$ is defined as an equivalent spherical radius, i.e. the
radius of the sphere having the same volume of the void. The discrepancy between observed and simulated profiles is addressed in Section 4.2.}
\label{profiles_all} 
\end{figure}

  \begin{figure*}
\vspace{-10 mm}
\includegraphics[trim=1.2cm 2cm 0.7cm 0.4cm, clip, width=1.9\columnwidth]{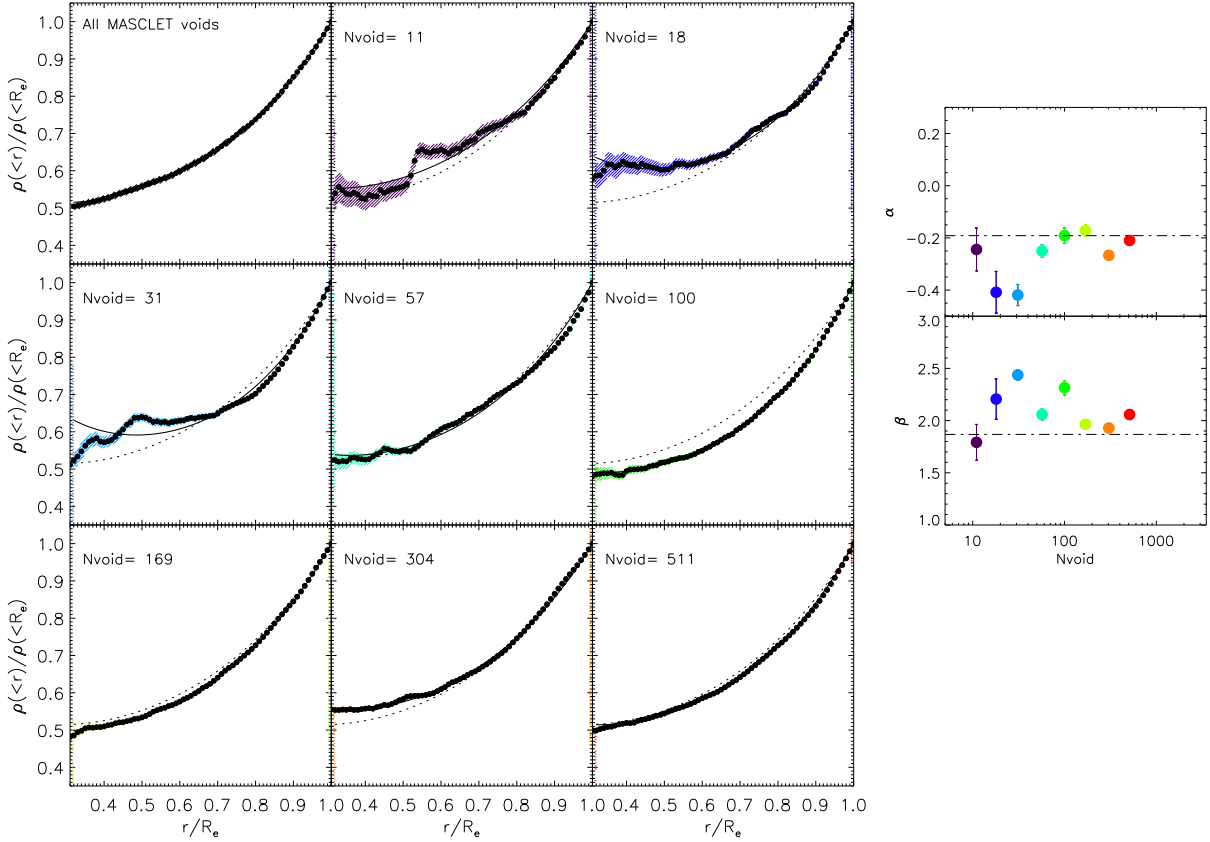}
\caption{Dependence of the simulated void density profiles on the
  number of voids used in the stacking. The profiles have been
  computed using the dark matter particles as density tracers. The different panels show the
  stacked profiles for subsamples of voids extracted from the original
  sample, populated with an increasing number of voids. 
  The black points indicate the radial value of the stacked void, colored shaded regions show the
  confidence regions determined by means of a bootstrapping, and the
  black solid line is the best-fit. The dotted black line reported in
  all the panels is the best-fit density profile of the stack drawn
  from the parent sample, when all the voids are included. The
  dependence of the best-fit parameters on the number of voids used in
  the stacking 
  is shown in the
  smaller panels on the right.  Error-bars have been computed with the aid
  of a bootstrapping, see text for further details (Section 4.1).
}
\label{profiles_nv} 
\end{figure*} 

To derive the void density profiles for the simulated voids, we can 
rely on two different tracers: the continuous density field and the dark
matter particles. In both cases, we compute the density in spherical
apertures, thus discarding the information about void shape. 
When using the continuous density field,  we  compute the profiles for
the individual voids and then use the bi-weight estimator at any
given aperture for getting the stacked profile. 
We have restricted the analysis of the profiles to 
$r \geqslant 0.3R_e$, as in the observed voids we can not reach regions at
smaller radii, due to
the paucity of galaxies.
In Fig. \ref{profiles_all}  we show the profile computed in this way
for all simulated voids larger than $7\,h^{-1}\,
Mpc$ (blue line).  

 The stacked profile is fitted by the two-parameters law proposed in
RQP13:
\begin{equation}\label{myfit}
\frac{\rho(<r)}{\rho_e}=\Big(\frac{r}{R_e}\Big)^{\alpha}\exp\Big[ \Big(\frac{r}{R_e}\Big)^{\beta}-1\Big]
\end{equation}
where $\rho(<r)$ is the density enclosed within the void-centric distance $r$,
$\rho_e$ is the density enclosed within the void effective radius
$R_e$ and 
$\alpha$ and $\beta$ are the best-fit parameters to be obtained
from the fit. 
We notice that in order to avoid a divergent profile for $r=0$, we
should require $\alpha, \beta \geqslant 0$. However, since we are applying
Eq. \ref{myfit} to a limited radial range ($0.3-1 R_e$), we allow
 $\alpha$ and $\beta$ to assume any value. 
This is particularly useful to
 quantify the behavior of the inner part of the profile in very
 different situations.  In fact, cases where
 $\alpha$ takes negative values do exist, as we show in the following sections. 
The best-fit parameters that we obtain for the stacked void of
Fig. \ref{profiles_all} are: $\alpha=0.06$
and $\beta=1.76$,  compatible with those determined in
RQP13, using a sample of smaller voids. 
We also show the profile when only the density of the baryonic
component is considered (orange
symbols).  As shown in RQP13, the
distribution of the gas within low density regions closely follows that
of dark matter. Indeed, the density profiles for the two components are
in a remarkable agreement and the best-fit parameters for the gas only
profile are: $\alpha=0.01$ and $\beta=1.65$.

\begin{figure*}
%\vspace{-10 mm}
\includegraphics[trim=1.2cm 2cm 0.7cm 0.4cm, clip, width=1.9\columnwidth]{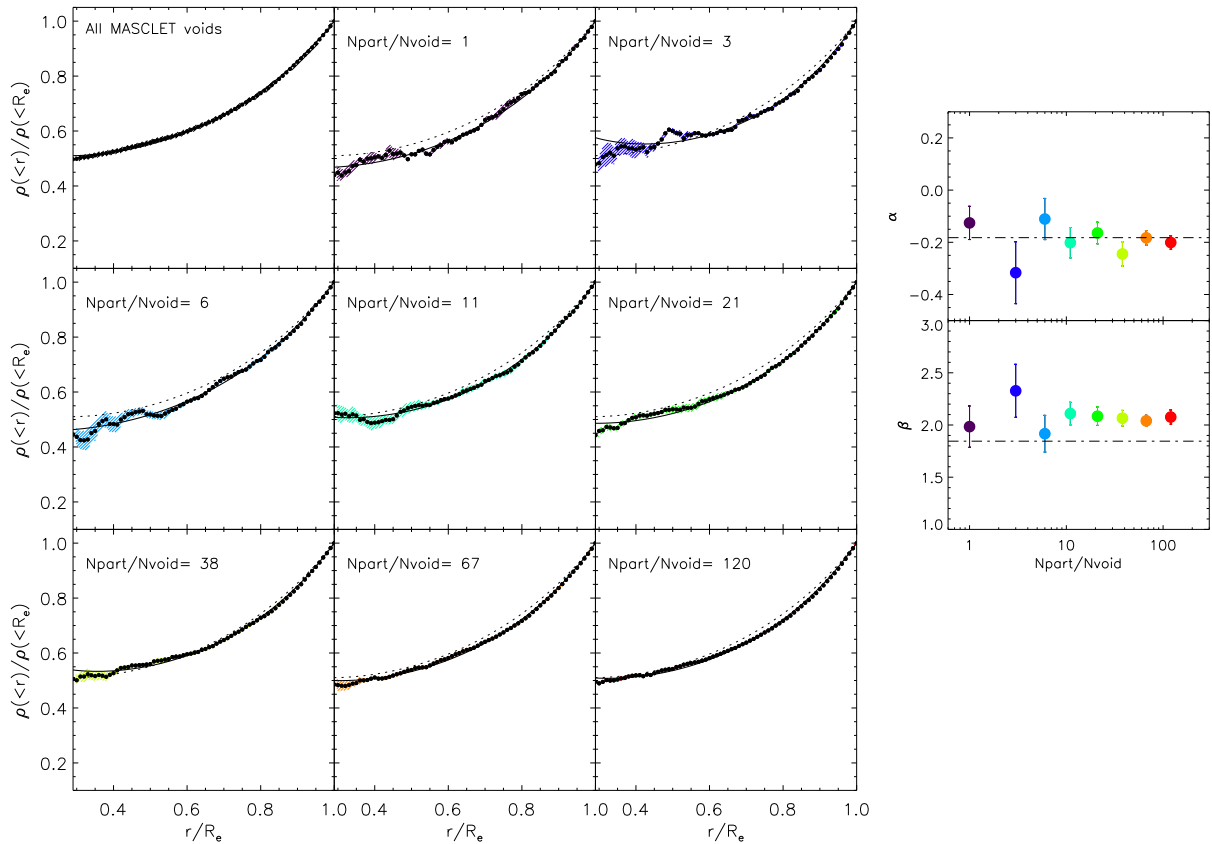}
\caption{Dependence of the simulated void density profiles
  on the average number of particles per void. The different panels show the
  stacked profiles for voids populated with an increasing number of
  particles, extracted from the original
  sample.  As in Fig. \ref{profiles_nv}, the black points indicate the radial value of the stacked void, colored shaded regions show the
  confidence regions, and the
  black solid line is the best-fit. The dotted black line reported in
  all the panels is the best-fit density profile of the stack drawn
  from the parent sample, when all void particles are included. The
  dependence of the best-fit parameters on the average number of
  particles per void   is shown in the
  smaller panels on the right.  
}
\label{profiles_ngv} 
\end{figure*} 

The second method  we use,  relies on the dark matter particles within
the void regions. In building the stack, we include, for each void, all
dark matter particles within the  radius limit, and rescale their
void-centric distance to the radius of the void hosting the particle. 
The particles are ranked according to their rescaled distance $r$,
from the smallest to the largest, and the stacked profile at $r$ is computed with the following expression:
\begin{equation}
\frac{\rho(<r)}{\rho_e}=\frac{1}{N_{void}}\sum_{i=1}^{N(<r)}\frac{m_iw_i}{(4/3)\pi(rR_{ei})^3\rho(<R_{ei})} 
\end{equation}
where the summation is intended over all the particles contained
within the rescaled radius $r$, $N_{void}$ is the total number of
voids entering in the stack, $m_i$ is the mass of the {\it i-th}
particle, $R_{ei}$ is the effective radius of the void containing the
{\it i-th} particle and $\rho(<R_{ei})$ is the density enclosed within
$R_{ei}$, computed with all the particles within
the void containing the {\it i-th} particle\footnote{in the
  computation of $\rho(<R_{ei})$ the particle masses are
  weighted according to  Eq. \ref{w1} and \ref{w2}.}.
The weights $w_i$ are intended to give  low
weights to the massive particles falling too close to the void
centers, which otherwise would bias the inner profile towards high values.
They are defined as:
\begin{equation}\label{w1}
w_i=(1-u^2)^2
\end{equation}
with:
\begin{equation}\label{w2}
u=min\left( \left|\frac{(\rho\rho(r)-\overline{\rho\rho}}{N_{\sigma}\sigma} \right|,1\right)
\end{equation} 
being $\rho\rho=\rho(r)/\rho(R_e)$ the density at the location $r$,
computed with the mass and location of the {\it i-th} particle, 
rescaled to the density at the void radius $\rho(R_e)$; 
$\overline{\rho\rho}$ is the median density
computed with 20 neighbour particles and $\sigma$ is the 
median absolute deviation, $N_{\sigma}$ is an adjustable parameter,
that in our configuration is chosen to be 6\footnote{This weighting
  scheme is the same adopted in the bi-weight estimator}.
This profile is shown by the green symbols in
Fig. \ref{profiles_all}. 
It slightly deviates from the one computed
with the continuous density field, because the density in the inner
part of voids can be seriously affected by the sparsity of the
particle distribution. The bi-weight estimator of the
individual profiles turns out to be a far more robust method when dealing with very
noisy data, such as the density in the very inner part
of voids. 
However,
with dark matter particles we
want to adopt the same method that can be used with the
SDSS voids, where individual profiles are difficult to be obtained,
given the paucity of galaxies living in them. 

To derive the void density profiles of the SDSS voids, we need to rely
on the luminous galaxies as density tracers.
We therefore adopt the same stacking procedure used for stacking the
dark matter particles in the simulated voids, by considering all the
void galaxies in the parent sample.
The observed void density profile is shown in
Fig. \ref{profiles_all} as black symbols. The functional form
expressed in Eq. \ref{myfit} turns out to be adequate in reproducing also
the observed profile, with $\alpha=0.50$ and $\beta=4.15$. We find
nevertheless that the observed profile is much steeper than the
simulated one.  Given the close agreement between gas and dark matter
density profiles, we can not ascribe such a steepness to a baryonic
bias. We investigate the origin of such difference in the following
sections.

\subsection{Impact of the undersampling}\label{undersampling}

\begin{figure*}
\includegraphics[trim=1.2cm 2cm 0.7cm 0.4cm, clip, width=1.9\columnwidth]{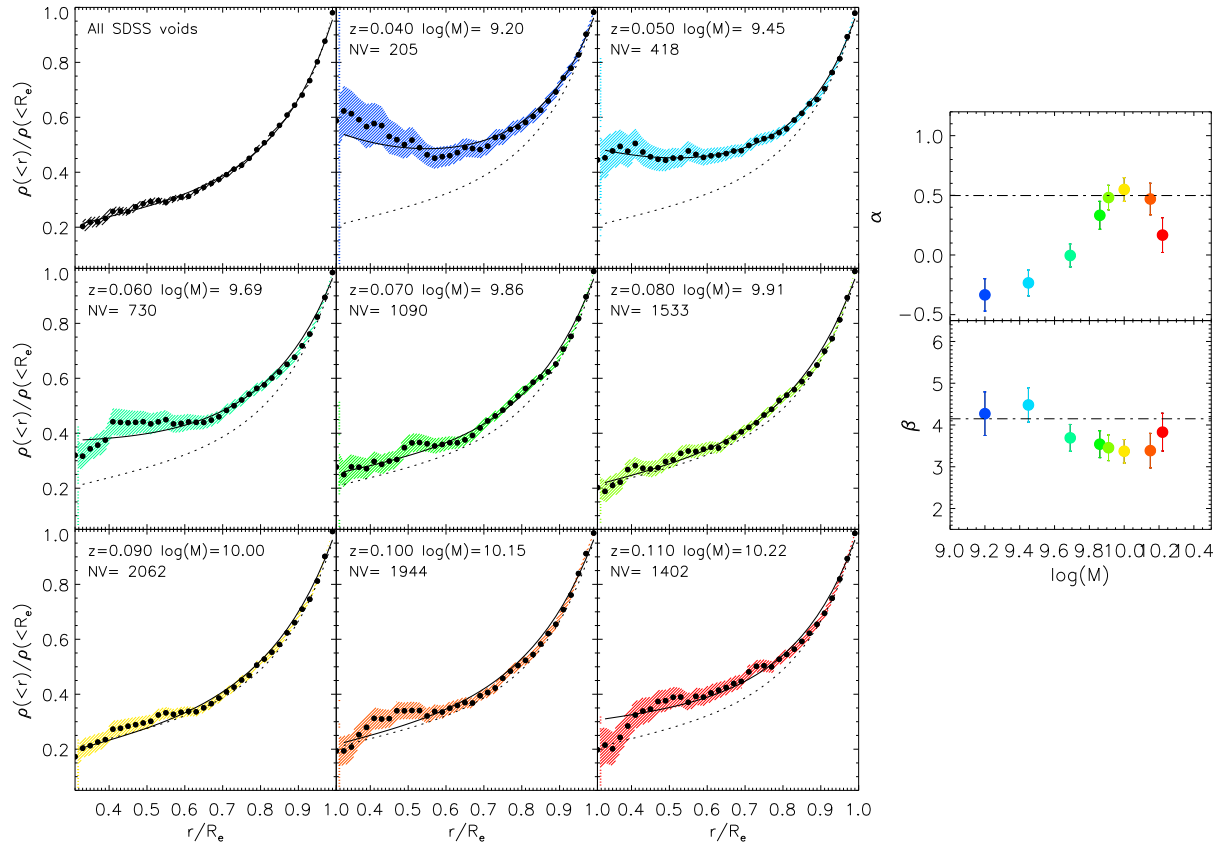}
\caption{Dependence of the observational void density profiles on the
  choice of the density tracers. Different panels show the density profiles
  for samples of voids lying at redshift below that indicated and
  using, as mass tracers, the galaxies more massive than the threshold
  mass at that redshift. The black points indicate the radial value of the stacked void, colored shaded regions show the
  confidence regions determined by means of a bootstrapping, and the
  black solid line is the best fit. The dotted black line reported in
  all the panels is the best-fit density profiles of the stack drawn
  from the parent sample (first panel). The
  dependence of the best-fit parameters on the threshold mass of the galaxies used 
  is shown in the smaller panels on the right.  
 }
\label{profiles_cuts} 
\end{figure*} 

A possible reason for the steepness of the observed density profile 
could lie in the sparsity of the galaxy distribution. The paucity of
galaxies within the observed voids, especially at small void-centric distances,
makes difficult to reconstruct the underlying distribution of matter. 
Indeed, in our catalogue of SDSS voids, the typical number of galaxies
populating the voids is 10, and several voids contain only 1-2
galaxies. Therefore, it is important to assess whether the limited statistics or
the low density of tracers (either dark matter particles in the
simulation or galaxies in the SDSS catalogue)  could
affect the shape of the void density profile.

In order to test the undersampling, we have used the simulated voids. 
The undersampling has been tested against both the number of voids
used in the stacking and the number of density tracers.
Since the resolution of the simulation used for this work is quite
modest, we do not find haloes and galaxies within the
voids. Therefore, to study the effect of the sparsity of the density tracers we rely
on the dark matter particles.

In assessing the effect of the limited statistics, we have randomly extracted subsamples of voids from the
parent catalogue. The resulting density profiles are shown in Fig. \ref{profiles_nv} for
subsamples populated with an increasing number of voids.
The dependence of the best-fit parameters on the
number of voids is shown in the small panels on the
right-hand side of Fig. \ref{profiles_nv}. The errors on $\alpha$ and $\beta$
have been estimated by means of a bootstrap resampling. For
each void subsample, we have generated 100 resamplings with
replacement and computed the stacked profile with the relative
best-fit values.  Their standard deviations give the errors on the
measured $\alpha$ and $\beta$.
The effect of the limited statistics is to increase the noise in
samples where very few voids are stacked, but no systematic effect is observed.
As shown by the best-fit parameter panels, $\alpha$ and $\beta$
converge to the reference values when at least 100 voids are stacked.

We have also analysed the undersampling effect by computing the
profiles with voids populated with an increasing number of dark matter
particles. The resulting profiles are shown in
Fig. \ref{profiles_ngv}.  The most important conclusion is that the paucity of tracers does not bias the
recovered profiles. As inferred from the behavior of the best-fit parameters of 
Fig. \ref{profiles_ngv}, voids populated with less than 10 particles
show some deviations from the reference values, because the profiles
are particularly noisy, but no systematic effect is observed.

To conclude, we can consider the void density profiles as reliable
when more than 100 voids are used in the stack and when they are
populated with at least 10 tracers.
In the SDSS stack void shown in Fig. \ref{profiles_all}, both conditions are satisfied, hence we can
not  ascribe the steepness of the observed profile to undersampling effects.

\subsection{Impact of the mass tracers}

In this section, we study the impact of the sample of galaxies
adopted on the resulting density profiles of the observed voids. 
This is particularly important if one wants to study the void density
profiles as a function of the radius of the voids. In fact, the
largest voids are more likely observed in the higher redshift bins, as
the sampled volume is larger. However, as a consequence of the
Malmquist bias, at high redshift only the brightest galaxies are observed (see 
 Fig. \ref{completeness}). Therefore, to compare voids located at
 different redshifts, we need to know whether the different tracers adopted can
 affect the resulting profile.

To do this, we have built volume limited samples of galaxies up to a given redshift and complete down to the corresponding
threshold mass limit. The redshift and mass limits are those
illustrated in Fig. \ref{completeness}. 
It is worth to emphasize that the choice of the galaxy sample used
only affects the recovered density profiles, leaving the sample of
voids unchanged. We do not consider samples
at $z<0.04$ as the number of voids is too limited and the profiles can
be affected by undersampling. 
In Fig. \ref{profiles_cuts} we show the recovered density profiles
using the different galaxy samples. The profiles appear to steepen as
galaxies at higher redshift and  higher stellar mass are
used. Interestingly, the profiles traced with the faintest galaxy samples
approach the simulated profiles shown in
Fig. \ref{profiles_all}.  
The steepening is particularly evident in the evolution of $\alpha$,
that becomes progressively higher as more massive galaxies are
concerned.  On the other hand, $\beta$ does not show any clear dependence
on the tracers, as the $\beta$ beta values are just scattered around the
reference values. 

\begin{figure}
\includegraphics[width=0.95\columnwidth]{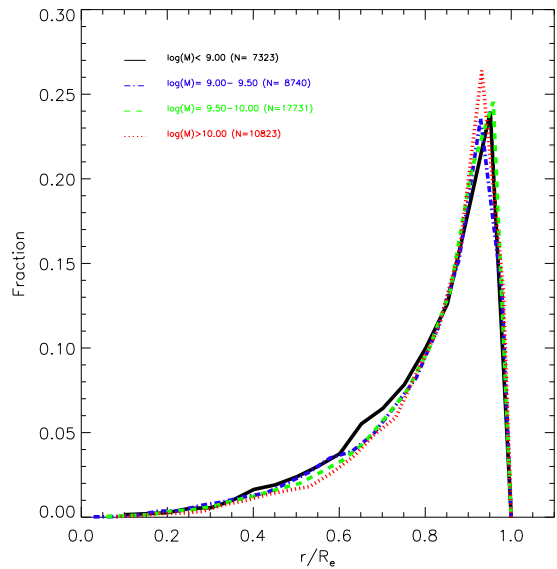}
\caption{Fraction of galaxies as a function of their distance to the
  void centre, normalized to the void radius. The different lines show
galaxies within different mass ranges, as indicated. }
\label{mass_segr} 
\end{figure}

We exclude redshift evolution as the reason for the profile
steepening observed in Fig. \ref{profiles_cuts}. Indeed, the evolution of voids in such a narrow redshift range
is expected to be negligible and should go in the opposite sense,
i.e. steeper profiles at lower redshift (see Figure 9 in RQP13). 
To understand how the choice of the mass tracers affects the
profiles, we show in Fig. \ref{mass_segr} the distribution of
void-centric distances of void galaxies of different masses. 
As expected, in all the samples the number of galaxies is extremely scarse in the inner
part and then it rapidly grows towards the edge, reaching a maximum at
$r/Re \sim 0.9$. Low mass galaxies appear to live closer to the void
centre, in particular there is an excess of dwarf galaxies at
$r/Re \sim 0.5$. Conversely, the massive galaxies are more
concentrated towards the void edge. A similar effect for the dwarf
systems has been also observed by \citet{Hoyle12}.
We argue that the steepness of the observed profile, with respect to
the simulated ones, can be explained
by the absence of tracers in the innermost regions of the observed
voids. It is not clear however whether such absence 
could be solved by using deeper data or it is just a consequence of
the galaxy bias. 
 
We note that a similar comparison of density profiles measured with 
different samples of galaxies has been shown by
\citet{Nadathur13}. They did not find any bias on the profile 
when considering tracers of different magnitude.
However, 
their galaxy samples are relatively bright ($Mr<-18.16+5log(h)$).
Void galaxies in our SDSS catalogue, are, by definition,
fainter than
$Mr=-20.17+5log(h)$, hence allowing us to probe the profiles using
also galaxies with very low mass. In fact, if only the highest mass 
bins were concerned,  
we would not observe such
dependence of the profile on the galaxy mass.

\subsection{Dependence on void radius}

\begin{figure*}
\includegraphics[trim=1.2cm 2cm 0.7cm 0.4cm, clip, width=1.9\columnwidth]{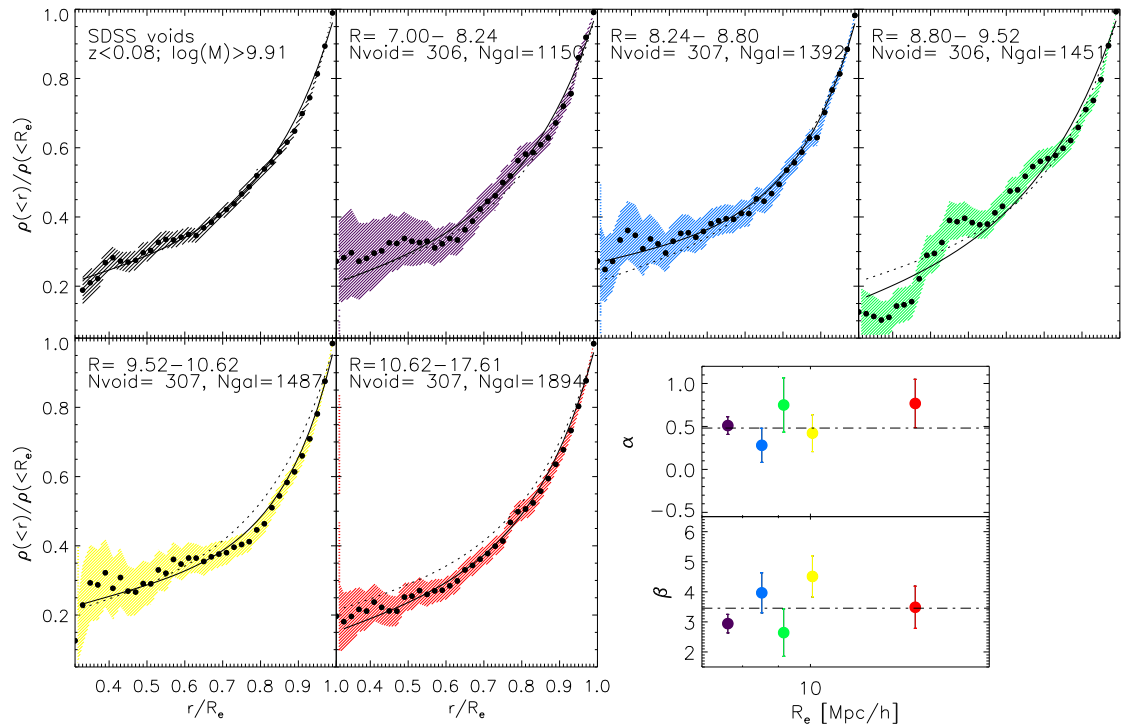}
\caption{Observational void density profiles as a function of void
  radius. The first panel shows the density profile measured from an
  homogenous sample of galaxies, with voids lying at $z<0.08$ and the
  mass tracers having stellar mass above $10^{9.9}\,M_{\odot}$. In the
  other panels, we show the density profiles of voids within different
  size bins. The black points indicate the radial value of the stacked void, colored shaded regions show the
  confidence regions determined by means of a bootstrapping, and the
  black solid line is the best fit. The dotted black line reported in
  all the panels is the best-fit density profile shown in the first
  panel. The lower-right panels show the dependence of the best-fit
  parameters $\alpha$ and $\beta$ on void radius.
 }
\label{profiles_rad} 
\end{figure*} 

\begin{figure*}
\includegraphics[trim=1.2cm 2cm 0.7cm 0.4cm, clip,width=1.9\columnwidth]{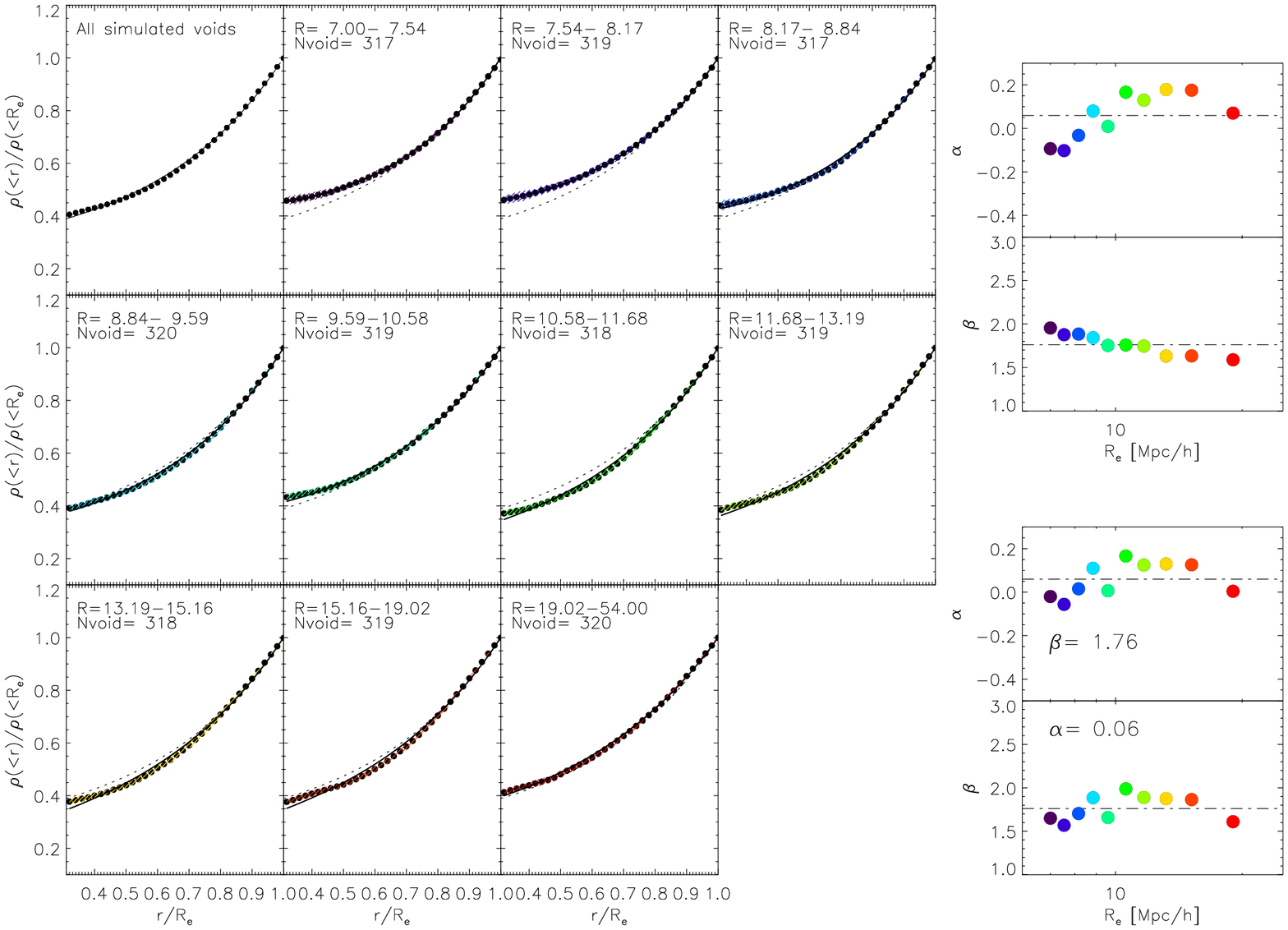}
\caption{Simulated void density profiles as a function of void
  radius. The first panel shows the mean density profile for all voids
  larger than $7\,h^{-1}Mpc$, whereas the other
  panels show the profiles for voids in different size intervals, as
  indicated.  In all panels black symbols indicate the bi-weight
  mean profile, the coloured shaded regions show the 1$\sigma$ confidence
  interval and the solid black line stays for the best-fit model. The best-fit profile of the parent sample shown in the
  first panel is reported, as dotted line, in all the other panels for reference.  The
  dependence of the best-fit parameters on void radius is shown in the
  smaller panels on the right, in the case of leaving both $\alpha$
  and $\beta$ free (upper panels) or fixing one of the two to the
  reference value (lower panels). 
 }
\label{profiles_rad_sim} 
\end{figure*} 

To assess the dependence of the void profiles on the void radius,
we use both the observed and simulated voids. 
In the observed voids, in order not to be affected by the bias
described in the previous section, we rely on a 
homogenous sample of galaxies.  We focus on voids
located at $z<0.08$ and use only galaxies more massive than the
threshold mass at this redshift ($10^{9.9}\,M_{\odot}$) for the stacking. We choose this
couple of redshift and mass because is the one maximizing the number
of galaxies (see lower panel of Fig. 1). We find 1539 voids and 7725
galaxies satisfying this criteria.

We divide the void sample in equi-populated subsamples, having
$\sim300$ voids each. This should limit the effect of noise at small
radii, shown in Section \ref{undersampling}.
The profiles for different void radii are shown in
Fig. \ref{profiles_rad}.
All the best-fit parameters, $\alpha$ and $\beta$, fall within 1-2
  $\sigma$ of the reference values, derived by fitting the profile of the parent
sample at $z<0.08$, without any dependence on the radius.
Indeed, the best-fit profile derived for the parent sample (dotted line) is
compatible with the profile shape in all the size bins.

We also probe the dependence of void profiles on radius by means of
the simulated voids. Fig. \ref{profiles_rad_sim} shows the density
profiles for different void sizes. The
density here refers to the total density, including dark matter and
gas. A similar result is obtained when using the dark matter particles
as density tracers. The void sample is divided in different size bins, containing
more than $300$ voids.
Given the large box of the simulation, we are able to probe even the largest
voids, having radii up to $\sim 50\,h^{-1}Mpc$. 
The profiles of each size bins are in good agreement with that of the
parent sample (dotted line). However, in the trend of the best-fit parameters  with radius (right-hand upper
panels), we observe a positive correlation of $\alpha$ with $R_e$ and
a negative correlation of $\beta$ with $R_e$.  This is due to
some degeneracy in the fit, as the two parameters  are not completely
independent. Therefore, we have fitted the two parameters
separately. Hence $\alpha$ ($\beta$) has been fitted by keeping
$\beta$ ($\alpha$) fixed and equal to its reference value. The results
are shown in the lower panels on the right-hand side of
Fig. \ref{profiles_rad_sim}.  In this case, we do not see any
correlation with $R_e$. We argue that the profile is independent on
the void radius.
 
We notice that  the void density profiles have also 
been tackled in RQP13, 
using an analogous simulation as the one presented in this
work, though with a much smaller volume. 
However, in that work the statistics of voids larger than
$8\,h^{-1}Mpc$ were too limited, due to the small volume of the
simulation, and it was not possible to draw robust conclusions. 

\section{Conclusions}\label{conclu}

We have robustly assessed the universality of void density profiles,
by means of a catalogue
of observed voids and a large cosmological simulation. 

The observed void catalogue has been drawn from the SDSS database, and includes spherical voids
whose radius is larger that  $7 \, h^{-1}\,  Mpc$ \citep{Varela12}. To
measure the density profiles in these  voids, we 
rely on the luminous galaxies.
As a matter of comparison, we have performed a large cosmological
simulation with the code MASCLET, devoted to follow the formation and
evolution of the low-density regions. This simulation has been
designed to target, with sufficient statistics, voids spanning a wide
range of radius. 
To this aim, we have simulated a large volume, having a comoving side
length of $512 \, h^{-1}\,  Mpc$,  with only one level of refinement
in the AMR grid, reaching the spatial resolution of
$0.5 \, h^{-1}\,  Mpc$. Since this  modest resolution does not allow to
follow the formation of structures in the simulated box, void galaxies
in our simulation do not form. Therefore, we adopt as density
tracer the continuous density field or, where a sparse distribution of
tracers is needed, the dark matter
particles within the void regions. 

The void density profiles recovered by means of the observed and
simulated voids share the same qualitative shape, showing a
significant underdensity in the centre and a sharp density increase 
approaching the void edges. Both profiles can be well described
by the functional form proposed in RQP13.  However,
the observed profile is significantly steeper than the simulated one.

To figure out the reasons for the steepness of the observed
profiles, we have assessed the impact of the number and type of
tracers on the resulting  density profile.
The sparsity of the density tracers has been investigated by means of
subsamplings of the simulated voids, populated with an increasing number of
particles. We have shown that even in the  less populated void samples, the
original density profile can be recovered with reasonable
accuracy. Stacks built with a limited number of voids or sparsely populated present
a significant noise at small radii, but no systematic effect with the
number of voids/tracers is observed.
The low impact of the sparsity of the tracers on the internal void
density profiles has been pointed out also by \citet{Sutter13},
using both dark matter particles and haloes as density tracers.

Nevertheless, we observe that the profile shape can have a significant
dependence on the type of galaxies used to trace the matter distribution. Within the
observed voids, the density
profiles recovered by means of faint samples of galaxies are 
shallower than those determined through the brighter galaxies. 
The reason for that lies in the galaxy mass segregation within
voids. In fact,  faint galaxies are those living closer to the
void centre and, thus, allow to probe the matter distribution even in
the innermost part of the voids. 

The strong impact of the type of galaxies chosen to trace the density, forces
us to use an homogenous sample of galaxies and voids, limited in
volume and magnitude, to assess the dependence of the void density
profile on the void radius. With such a sample, we have
demonstrated the insensitivity of the observed void profile on void radius. 
Likewise, by using our simulated sample of voids, we do not observe
any dependence of the profile shape on the void size, and the same
best-fit can correctly describe voids whose size ranges from 7 to   
$\sim 50 \, h^{-1}\,  Mpc$.

Finally, we note that the difference in profile between the observed
and simulated voids can not be driven by the different algorithms used
to identify voids. Indeed, 
the density profile
of our SDSS stack is very similar to the profile published in Pan et
al. (2012), using the same SDSS DR7 dataset, albeit with a 
completely different void finder.
Moreover, our simulated void
density profiles are in remarkable agreement with the simulations of
\citet{Colberg05}, where voids are identified through spherical underdensities.
Therefore, we argue that the difference between observed and simulated
void density profiles is a robust result and is 
due to the biased tracers used, when relying
on the observed galaxies. 
To corroborate this hypothesis, we definitely need high resolution
simulations, capable to follow structure formation in the most
rarefied regions of the Universe.

\section*{Acknowledgements} 

We  are grateful to Ignacio Trujillo for useful discussions. This work was  supported by the Spanish Ministerio de Econom\'{\i}a y Competitividad 
(MINECO, grants   AYA2010-21322-C03-01) and    the   Generalitat   Valenciana   (grant
PROMETEO-2009-103). J.V. did part of the work thanks to a post-doc fellowship from the former Spanish Ministry of Science and Innovation under programs 3I2005 and 3I2406. J.V. also acknowledges the financial support from the FITE (Fondos de Inversión de Teruel) and the Spanish grant AYA2012-30789.


\begin{thebibliography}{}

\bibitem[\protect\citeauthoryear{Alcock 
\& Paczynski}{1979}]{AP79} Alcock C., Paczynski B., 1979, Nature, 281, 358 

\bibitem[\protect\citeauthoryear{Aragon-Calvo 
\& Szalay}{2013}]{AragonCalvo13} Aragon-Calvo M.~A., Szalay A.~S., 2013, MNRAS, 428, 3409 

\bibitem[\protect\citeauthoryear{Blanton et 
al.}{2005}]{Blanton05} Blanton M.~R., et al., 2005, AJ, 129, 2562 

\bibitem[\protect\citeauthoryear{Blanton 
\& Roweis}{2007}]{Blanton07} Blanton M.~R., Roweis S., 2007, AJ, 133, 734 

\bibitem[\protect\citeauthoryear{Bond, Kofman, 
\& Pogosyan}{1996}]{Bond96} Bond J.~R., Kofman L., Pogosyan D., 1996,
Nature, 380, 603

\bibitem[\protect\citeauthoryear{Bos et al.}{2012}]{Bos12} 
Bos E.~G.~P., van de Weygaert R., Dolag K., Pettorino V., 2012, MNRAS, 426, 
440 

\bibitem[\protect\citeauthoryear{Park 
\& Lee}{2007}]{2007PhRvL..98h1301P} Park D., Lee J., 2007, PhRvL, 98, 081301 

\bibitem[\protect\citeauthoryear{Colberg et 
al.}{2008}]{Colberg08} Colberg J.~M., et al., 2008, MNRAS, 387, 
933 

\bibitem[\protect\citeauthoryear{Colberg et 
al.}{2005}]{Colberg05} Colberg J.~M., Sheth R.~K., Diaferio A., 
Gao L., Yoshida N., 2005, MNRAS, 360, 216 

\bibitem[\protect\citeauthoryear{Colless et 
al.}{2001}]{Colless01} Colless M., et al., 2001, MNRAS, 328, 1039 

\bibitem[\protect\citeauthoryear{Eisenstein \& Hu}{1998}]
{EiHu98} Eisenstein D.J., Hu W., 1998, ApJ, 511, 5

\bibitem[\protect\citeauthoryear{Gregory 
\& Thompson}{1978}]{Gregory78} Gregory S.~A., Thompson L.~A., 1978, ApJ, 222, 784 

\bibitem[\protect\citeauthoryear{Gottl{\"o}ber et 
al.}{2003}]{Gottlober03} Gottl{\"o}ber S., {\L}okas E.~L., Klypin 
A., Hoffman Y., 2003, MNRAS, 344, 715 

\bibitem[\protect\citeauthoryear{Hoyle, Vogeley, 
\& Pan}{2012}]{Hoyle12} Hoyle F., Vogeley M.~S., Pan D., 2012, MNRAS, 426, 3041 

\bibitem[\protect\citeauthoryear{Kirshner et 
al.}{1981}]{Kirshner81} Kirshner R.~P., Oemler A., Jr., Schechter 
P.~L., Shectman S.~A., 1981, ApJ, 248, L57 

\bibitem[\protect\citeauthoryear{Krause et al.}{2013}]{Krause13} 
Krause E., Chang T.-C., Dor{\'e} O., Umetsu K., 2013, ApJ, 762, L20 

\bibitem[\protect\citeauthoryear{Lavaux 
\& Wandelt}{2010}]{Lavaux10} Lavaux G., Wandelt B.~D., 2010, MNRAS, 403, 1392 

\bibitem[\protect\citeauthoryear{Lavaux 
\& Wandelt}{2012}]{Lavaux12} Lavaux G., Wandelt B.~D., 2012, ApJ, 754, 109 

\bibitem[\protect\citeauthoryear{Laureijs et 
al.}{2011}]{Laureijs11} Laureijs R., et al., 2011, arXiv, 
arXiv:1110.3193 

\bibitem[\protect\citeauthoryear{Melchior et 
al.}{2013}]{Melchior13} Melchior P., Sutter P.~M., Sheldon E.~S., 
Krause E., Wandelt B.~D., 2013, arXiv, arXiv:1309.2045 

\bibitem[\protect\citeauthoryear{Nadathur 
\& Hotchkiss}{2013}]{Nadathur13} Nadathur S., Hotchkiss S., 2013, arXiv, arXiv:1310.2791 

\bibitem[\protect\citeauthoryear{Neyrinck}{2008}]{Neyrinck08} 
Neyrinck M.~C., 2008, MNRAS, 386, 2101 

\bibitem[\protect\citeauthoryear{Pan et al.}{2012}]{Pan12} 
Pan D.~C., Vogeley M.~S., Hoyle F., Choi Y.-Y., Park C., 2012, MNRAS, 421, 
926 

\bibitem[\protect\citeauthoryear{Park 
\& Lee}{2007}]{Park07} Park D., Lee J., 2007, PhRvL, 98, 081301 

\bibitem[\protect\citeauthoryear{Patiri et al.}{2006}]{Patiri06b} 
Patiri S.~G., Prada F., Holtzman J., Klypin A., Betancort-Rijo J., 2006b, 
MNRAS, 372, 1710

\bibitem[\protect\citeauthoryear{Platen, van de Weygaert, 
\& Jones}{2007}]{Platen07} Platen E., van de Weygaert R., Jones
B.~J.~T., 2007, MNRAS, 380, 551 

\bibitem[\protect\citeauthoryear{Plionis 
\& Basilakos}{2002}]{Plionis02} Plionis M., Basilakos S., 2002, MNRAS,
330, 399 

\bibitem[\protect\citeauthoryear{Quilis}{2004}]{Quilis04} Quilis 
V., 2004, MNRAS, 352, 1426 


\bibitem[\protect\citeauthoryear{Ricciardelli, Quilis, 
\& Planelles}{2013}]{Ricciardelli13} Ricciardelli E., Quilis V.,
Planelles S., 2013, MNRAS, 434, 1192  


\bibitem[\protect\citeauthoryear{Sheth 
\& van de Weygaert}{2004}]{Sheth04} Sheth R.~K., van de Weygaert R., 2004, MNRAS, 350, 517 

\bibitem[\protect\citeauthoryear{Sutter et al.}{2012}]{Sutter12a} 
Sutter P.~M., Lavaux G., Wandelt B.~D., Weinberg D.~H., 2012a, ApJ, 761, 187 


\bibitem[\protect\citeauthoryear{Sutter et al.}{2012}]{Sutter12b} 
Sutter P.~M., Lavaux G., Wandelt B.~D., Weinberg D.~H., 2012b, ApJ, 761, 44 

\bibitem[\protect\citeauthoryear{Sutter et al.}{2013}]{Sutter13} 
Sutter P.~M., Lavaux G., Wandelt B.~D., Hamaus N., Weinberg D.~H., Warren 
M.~S., 2013, arXiv, arXiv:1309.5087 



\bibitem[\protect\citeauthoryear{York et al.}{2000}]{York00} 
York D.~G., et al., 2000, AJ, 120, 1579 

\bibitem[\protect\citeauthoryear{van de Weygaert 
\& Platen}{2011}]{Vandew11} van de Weygaert R., Platen E., 2011, IJMPS, 1, 41 


\bibitem[\protect\citeauthoryear{Varela et al.}{2012}]{Varela12} 
Varela J., Betancort-Rijo J., Trujillo I., Ricciardelli E., 2012, ApJ, 744, 
82



\end{thebibliography}
\end{document}